\documentclass[prd,showpacs,showkeys,secnumarabic,12pt]{revtex4}
\usepackage{amsmath}

\newcommand{\pni}{\par\noindent}
\begin{document}
\title{On the classical confinement of  test particles to a 
thin 3-brane in the absence of non-gravitational forces} 
\author {M. La Camera} \email{lacamera@ge.infn.it} \affiliation 
{Department of Physics and INFN - University of Genoa\\Via 
Dodecaneso 33, 16146 Genoa, Italy} 
%\date{} 
\begin{abstract}
The classical confinement condition of test  particles to a brane
universe in the absence of non-gravitational forces is 
transformed using the Hamilton-Jacobi formalism. The transformed 
condition provides a direct criterion  for selecting in a 
cosmological scenario 5D bulk manifolds wherein it is possible to
obtain  confinement of trajectories to 4D hypersurfaces  purely 
due to classical gravitational effects. 
\end{abstract} 
\pacs{04.50.-h, 04.20.Cv} 
\keywords{Brane Theory; Classical Confinement } 
\maketitle  
\section{I\lowercase{ntroduction}}
In the braneworld theories with non-compact extra dimensions it 
is postulated that particles and fields of the standard model are
confined to the brane universe, while the graviton is assumed to 
propagate both in the bulk and in the brane. Within the classical
framework of this scenario, confining a test particle to the 
brane eliminates the effects of extra dimensions rendering them 
undetectable. In general  non-gravitational forces acting in the 
bulk and orthogonal to spacetime are needed in order to keep the 
test particles moving on the brane, the source of these confining
forces being interpreted in different manners.  If the notion of 
confinement must appear in any reasonable theory with non-compact
extra dimensions non-gravitational forces can not be excluded 
\textit{a priori}. Confinement due to oscillatory behaviour has 
been proved in five-dimensional relativity with two times 
${}^{1,2}$. Indeed, null paths of massless particles in 5D 
geodesic motion can appear in 4D as timelike paths of massive 
particles which undergo oscillations in the 5D dimension around 
the 4D hypersurface. Moreover in the four-dimensional spacetime 
there appears a non-gravitational force which gives rise to the 
possibility of a cosmological variation of the rest masses of the
particles with consequent departure from geodesic motion 
. However it may worth studying under which 
conditions confinement is possible without the introduction of 
force fields, other than gravity, living in the bulk. Employing a
phase space analysis of the splitting from the 5D to the 4D 
geodesic motion  Dahia, Romero, da Silva  and  Tavakol ${}^{3,4}$
investigated the possible confinement of particles and 
photons in the neighbourhood of a four-dimensional hypersurface 
in five-dimensional warped product spaces and found a form of 
toroidal confinement. This quasi-confinement, which is 
oscillatory and neutrally stable, is  due to classical  
gravitational effects without requiring the presence of other 
non-gravitational mechanisms. Our approach differs in various 
aspects from the previous one, but we find the similar result,
that is, if the bulk geometry satisfies some general conditions 
confinement of  test particles to a thin 3-brane is possible 
without the presence of non-gravitational mechanisms. In this 
paper we shall consider a foliation of the bulk, each leaf of the
foliation representing a brane, and we shall succeed in  
selecting 5D bulk metrics which allow to choose at least one 
brane of the foliation where confinement is possible without the 
introduction of non-gravitational forces.   We shall deal with a 
five-dimensional bulk and the geodesic motion of test particles 
will happen in a 5D background with three-dimensional isotropy 
and homogeneity. The dynamics of test  particles as observed in 
4D is generally discussed on the basis of the splitting of the 
geodesic equation in 5D. As pointed out by Ponce de Leon  
${}^{5,6}$ this process has some drawbacks with regard to the 
definition of the so called ``fifth force''. To overcome these 
drawbacks Ponce de Leon has analized the 5D geodesic equation in 
terms of local basis vectors and, suitably redefining various 
quantities, obtained a more correct description of test particles
trajectories.  An alternative formalism was presented
by Seahra and Wesson ${}^{7-9}$ who based their analysis on a 
covariant foliation of the 5D manifold using 3+1 spacetime slices
orthogonal to the extra dimension and derived the form of the 
classical non-gravitational force required to confine particles 
to a 4D hypersurface. We shall start from these last results  
but, as suggested by Ponce de Leon ${}^{10}$ in a somewhat 
different context, we shall go on utilizing the Hamilton-Jacobi 
(H-J) method. Indeed, the H-J formalism, where one has to deal 
with a ``scalar'' equation, will prove to be adequate to study 
the geodesic motion of test particles. As a final result it will 
possible to select, in a cosmological scenario, 5D bulk manifolds
wherein the motion of test particles can be confined at least to 
one 4D hypersurface of the foliation  without the introduction of
non-gravitational forces. In all the others 4D hypersurfaces of 
the foliation either the confinement is guaranted by 
non-gravitational forces or, in their absence, an observer on the
brane will perceive that test particles move subject to an extra 
force and with variable 4D rest mass, as described by a number of
authors in the literature ${}^{11-23}$. \pni 
\textit{Conventions}. Throughout the paper the 5D metric 
signature is taken to be $(+,+,+,-,\varepsilon)$ where 
$\varepsilon$ can be $+1$ or $-1$ depending on whether the extra 
dimension is spacelike or timelike, while the choice of 4D metric
signature is $(+,+,+,-)$. The spacetimes coordinates are labelled
$x^i=(r,\vartheta,\varphi), x^4=t$. The extra coordinate is 
$x^5=y$. Bulk indices will be denoted by capital Latin letters 
and brane  indices by lower Greek letters. 
\section{C\lowercase{onfinement of 5}D \lowercase{trajectories to
4}D \lowercase{hypersurfaces}}
The study of higher  dimensional particle motion will be 
performed in the foliating approach. In this section we shall 
give a concise description of the geometric construction and of 
the confinement condition obtained by Seahra and  Wesson   
${}^{7-9}$. These authors  considered a 5D manifold $M$ described
by the metric $g_{AB}(x^A)$ and introduced a scalar function 
$\ell=\ell(x^A)$ which defines a foliation of this  manifold by a
series of 4D hypersurfaces $l$ = constant denoted by 
$\Sigma_\ell$. The 5D manifold was referred to as ``the bulk'' 
so each leaf of the foliation is ``a brane''. The hypersurfaces 
$\Sigma_\ell$ were assumed to have a normal vector given by 
\begin{equation} n^A = \varepsilon\,\Phi\,\partial_A\ell,  
\quad n_A n^A = \varepsilon 
\end{equation}
The scalar $\Phi$ which normalizes $n^A$ is known as the lapse
function. The projector tensor  $h_{AB}$ from the bulk to the 
hypersurfaces is 
\begin{equation}
h_{AB}=g_{AB} - \varepsilon\, n_A n_B
\end{equation}
This tensor is symmetric and orthogonal to $n^A$.
Each hypersurface $\Sigma_\ell$ was mapped by a 4D coordinate 
system $\{\tilde{x}^\alpha\}$. The four basis vectors
\begin{equation}
e^A_\alpha = \dfrac{\partial x^A}{\partial\tilde{x}^\alpha}\, 
\qquad \mathrm{with} \qquad n_A e^A_\alpha = 0
\end{equation}
are tangent to the $\Sigma_\ell$ hypersurfaces and orthogonal to 
$n^A$. These basis vectors can be used to project 5D objects 
onto $\Sigma_\ell$ hypersurfaces.  The induced metric on the 
$\Sigma_\ell$ hypersurfaces is given by
\begin{equation}
h_{\alpha\beta} = e^A_\alpha e^B_\beta g_{AB} =  e^A_\alpha 
e^B_\beta h_{AB}
\end{equation}
Clearly $\{\tilde{x}^\alpha,\ell\}$ defines an  alternative  
coordinate system to $\{x^\alpha,y\}$ on $M$. 5D vectors were 
decomposed into the sum of a part tangent  to $\Sigma_\ell$ and a
part normal to $\Sigma_\ell$. For $dx^A$ it results 
\begin{equation}
dx^A = e^A_\alpha d\tilde{x}^\alpha + \left(N^\alpha 
e^A_\alpha +\Phi\, n^A \right) d\ell 
\end{equation}
The 4D vector $N^\alpha$ is called the  shift vector and it 
describes how the $\{\tilde{x}^\alpha\}$ coordinate system 
changes as one moves from a given  $\Sigma_\ell$ hypersurface to 
another. The 5D line element was then  written as
\begin{widetext}
\begin{equation} 
ds_5^2 = h_{\alpha\beta}\left(d\tilde{x}^\alpha+N^\alpha 
d\ell\right)\left(d\tilde{x}^\beta+N^\beta d\ell\right) 
+\varepsilon \,\Phi^2\,d\ell^2 
\end{equation}
\end{widetext}
Finally we recall how the  confinement condition    was 
obtained in ${}^{7-9}$. The equations of motion for a test  
particle moving in the bulk were given by
\begin{equation}
u^B\,\nabla_B\,u^A = \mathcal{F}^A, \quad u^A = 
\dfrac{d\tilde{x}^A}{d\lambda} 
\end{equation}
where $\mathcal{F}$ is some non-gravitational force per unit mass
and $\lambda$ is a 5D affine parameter.  These equations  
were decomposed  into relations involving the particle velocity 
$u^\alpha =  e_A^\alpha\, u^A$ tangent to the $\Sigma_\ell$ 
foliation  and the particle velocity $\xi = n_A\,u^A = 
\varepsilon\,\Phi \,\dfrac{d\ell}{d\lambda}$ parallel to the 
normal direction . In particular, the acceleration along the 
normal direction was found to be 
\begin{equation} 
u^A\,\nabla_A\,\xi = K_{\alpha\beta}\,u^\alpha\,u^\beta -\, 
\xi\,n^A\,u^B\,\nabla_A\,n_B + \mathcal{F}_n 
\end{equation}
where $\mathcal{F}_n = n_A \mathcal{F}^A$ and 
$K_{\alpha\beta}$ is the extrinsic curvature of the 
hypersurfaces $\ell$ = constant:
\begin{equation} 
K_{\alpha\beta} =  e_\alpha^A e_\beta^B \nabla_A n_B
\end{equation} 
Now, if a test particle is confined to a given 
$\Sigma_\ell$ hypersurface, its $\ell$ coordinate must be 
constant. This implies that velocity and acceleration along the 
normal direction must vanish so eq. (8) with $\xi=0$ yields 
\begin{equation}
K_{\alpha\beta}\,u^\alpha\,u^\beta + \mathcal{F}_n = 0
\end{equation}
which is the confinement condition in presence of 
non-gravitational forces obtained in ${}^{7-9}$. In this work 
instead we shall require classical confinement due to 
gravitational forces alone, so eq. (10)  will change to
\begin{equation}
K_{\alpha\beta}\,u^\alpha\,u^\beta = 0
\end{equation}
As discussed in ${}^7$, this is a necessary condition for 
confinement which is a bilinear combinations between the 
components of the four-velocity and does not imply 
$K_{\alpha\beta} = 0$ in general. If a member $\Sigma_\ell$ of 
the foliation satisfies $K_{\alpha\beta} = 0$, which is termed 
the totally geodesic condition, then geodesic on that 4D 
hypersurface are also geodesic of the 5D manifold $M$ ${}^{24}$. 
In this case, as pointed  out by Wesson ${}^{25,26}$, the weak 
equivalence principle in 4D can be understood as a geometrical 
symmetry of 5D. Vice versa, if all the geodesics in  
$\Sigma_\ell$ are also geodesic in $M$, then  $K_{\alpha\beta}$ 
is necessarily zero. Coming back to the purpose of the present 
paper, the constraint $K_{\alpha\beta}\,u^\alpha\,u^\beta = 0$ 
requires the knowledge of the four-velocity $u^\alpha$ which can 
be obtained, apart from mathematical difficulties, either from 
the spacetime components of the 5D geodesic equation or, as we 
shall do in this paper, from the H-J equation. If the constraint 
can be solved on a particular $\Sigma_\ell$ hypersurface it seems
reasonable, as conjectured also in ${}^{10}$, to choose this 
hypersurface as the correct representation of our 
four-dimensional spacetime. If the constraint can not be solved 
an observer on that hypersurface will perceive the test particles
move under the influence of an extra  force with their 4D masses 
variable in time. In the above case with no confinement, if one 
adopts the most conservative point of view that confinement is a 
prerequisite for any reasonable theory with non-compact extra 
dimensions, then he has to introduce a non-gravitational 
confining force per unit mass given  by $\mathcal{F}^A = - 
(K_{\alpha\beta}\,u^\alpha\, u^\beta)\,n^A $. In the next section
we shall give a procedure for selecting 5D bulk manifolds wherein
it is possible to achieve confinement in the absence of 
non-gravitational forces. 
\section{C\lowercase{onfinement 
condition in the} H\lowercase{amilton}-J\lowercase{acobi 
formalism}} 
Let us now return to the line element (6) and make some choices 
which will result in significant simplification of the following 
formulae. In braneworld theory the induced metric  $h_{\alpha\beta}$ is 
commonly identified with the spacetime metric  $g_{\alpha\beta}$.
In this paper we shall follow this approach and choose 
$\tilde{x}^\alpha = x^\alpha$. Moreover we shall  select a 
``stationary'' 4D coordinate frame setting $N^\alpha =0$ and let 
the other foliation parameter $\Phi$ depend on the coordinates. 
In the cosmological scenario that we  consider,  our 
homogeneous and isotropic  universe is embedded in a 
higher-dimensional manifold, so the 5D line element (6) will be 
rewritten in the usual form as
\begin{equation} ds_5^2 = a^2(t,\ell)\,d\sigma^2 - 
n^2(t,\ell)\,dt^2 + \varepsilon\,\Phi^2(t,\ell)\,d\ell^2 
%\hspace{1cm} 
\end{equation} 
where
\begin{equation}
d\sigma^2 = dr^2 + r^2\,(d\vartheta^2+\sin^2\vartheta d\varphi^2)
\end{equation}
$r,\vartheta, \varphi$ and $t$ are the  coordinates for  a 
spacetime with spherically symmetric spatial sections while 
$\ell=\ell(x^A)$ is the scalar function which defines the 
foliation of the 5D manifold. To solve the confinement condition 
(11) we shall not use the spacetime components of the 5D geodesic
equation but, as suggested in a somewhat different context 
${}^{10}$, we shall consider the bulk geodesic motion of massive 
test particles by means of the Hamilton-Jacobi equation which in 
5D is given by 
\begin{equation} g^{AB}\,\left(\dfrac{\partial 
S}{\partial x^A}\right)\, \left(\dfrac{\partial S}{\partial 
x^B}\right) =-\, m_5^2 
\end{equation}
where $m_5^2$  is the 5D rest mass of the particle and 
$S(x^\alpha,\ell)$ is the five-dimensional action related to its 
5D momentum by 
\begin{equation}
P^A = g^{AB}\,\left(\dfrac{\partial S} {\partial x^B}\right) 
\end{equation}
Because of  our choices on the spacetime metric, the 4D 
components of $P^A$ are already projected onto a brane. We 
identify the affine parameter $\lambda$ in (7) with the 5D proper
time $\tau_5$ and  write $u^\alpha = P^\alpha/m_5$ so in order to
obtain the four-velocity  we have  to know the action 
$S(x^\alpha,\ell)$. In the case of massless particles the 
trajectory in 5D is along isotropic geodesics and is given by the
Eikonal equation which can be obtained from the above formulae by
setting $m_5^2=0$ in the H-J equation (14) and substituting the 
momentum $P^A$ in eq. (15) with the 5D wave vector $k^A$. In the 
massless case it will be therefore used $k^\alpha$ instead of the
velocity $u^\alpha$.  In the spacetime with spherical simmetry 
which we  consider, test particles move on a plane passing 
through the center. We take this plane as the $\vartheta = \pi/2$
plane. Thus the H-J equation for the metric (12) is 
\begin{widetext}\begin{equation} 
\dfrac{1}{a^2}\,\left[\left(\dfrac{\partial 
S}{\partial r}\right)^2 + \dfrac{1}{r^2}\,\left(\dfrac{\partial 
S}{\partial \varphi} \right)^2\right]-\dfrac{1}{n^2}\, 
\left(\dfrac{\partial S}{\partial t}\right)^2 
+\dfrac{1}{\varepsilon\,\Phi^2}\, \left(\dfrac{\partial 
S}{\partial \ell}\right)^2 = -\,m_5^2 
\end{equation} \end{widetext}
It is worth noticing that in the  4D proper time  $\tau_4$ 
parametrization ${}^7$ the four-velocity is $v^\alpha =
d x^\alpha/d\tau_4$ and  the 4D momentum becomes $P^\alpha =
m_4 v^\alpha$ where $m_4$ is the 4D rest mass. Therefore the H-J 
equation (16) gives the following relation  between the rest mass
in 4D and in 5D
\begin{equation}
m_4^2 = m_5^2 + \dfrac{1}{\varepsilon\,\Phi^2}\, 
\left(\dfrac{\partial S}{\partial \ell}\right)^2 
\end{equation}
so the rest mass of a particle as perceived by an observer in 4D 
varies as a result of the 5D motion along the extra dimension, a 
result  known in the literature that we shall obtain again below.
The action $S(x^\alpha,\ell)$ in (16) separates as 
\begin{equation}
S(x^\alpha,\ell) = S_r(r)+L\,\varphi+S_{t\ell}(t,\ell)
\end{equation}
where $L$ is the angular momentum. Putting
\begin{equation}
\left(\dfrac{\partial S_r}{\partial r}\right)^2 +\dfrac{L^2}{r^2}
= C^2 \geq 0
\end{equation}
where $C^2$ is a separation constant related to the motion in 
space, we finally obtain
\begin{equation}
\dfrac{1}{n^2}\, \left(\dfrac{\partial 
S_{t\ell}}{\partial t}\right)^2 
-\,\dfrac{1}{\varepsilon\,\Phi^2}\, \left(\dfrac{\partial 
S_{t\ell}}{\partial \ell}\right)^2 = \,m_5^2 + \dfrac{C^2}{a^2}
\end{equation}
Remembering (15), the particle velocity normal to  $\Sigma_\ell$ 
can be written as 
\begin{equation}
\xi = \varepsilon\,\Phi\,\dfrac{d\ell}{d\tau_5} = 
\dfrac{1}{m_5\,\Phi}\,\dfrac{\partial S_{t\ell}}{\partial \ell}
\end{equation}
and the acceleration (8) becomes
\begin{widetext}
\begin{equation}
u^A\,\nabla_A\,\xi =
\dfrac{1}{m_5^2\,\Phi}\left[\dfrac{C^2}{a^3}\dfrac{\partial 
a}{\partial \ell}-\, \dfrac{1}{n^3}\dfrac{\partial  n}{\partial 
\ell}\,\left(\dfrac{\partial S_{t\ell}}{\partial t}\right)^2 
\right] + \,\dfrac{\xi}{m_5\,n^2}\, \left(\dfrac{\partial 
S_{t\ell}}{\partial t}\right)\dfrac{\partial \ln{\Phi}}{\partial 
t} \end{equation}
\end{widetext}
We shall require that confinement becomes  possible on a 
hypersurface $\Sigma_0$ corresponding to $\ell = \ell_0$ with  
$a, n$ and  $\Phi$ finite and non-zero on this hypersurface. This
implies that the velocity and the acceleration given, 
respectively, in (21) and (22) must vanish on $\Sigma_0$ so we 
get 
\begin{subequations} \begin{eqnarray}
\left(\dfrac{1}{\Phi}\,\dfrac{\partial S_{t\ell}}{\partial 
\ell}\right)_{\ell=\ell_0} = 0 \label{equation1} \\ 
\left[\dfrac{C^2}{a^3}\,\dfrac{\partial a}{\partial \ell} -\,
\dfrac{1}{n^3}\,\dfrac{\partial n}{\partial \ell}\, 
\left(\dfrac{\partial S_{t\ell}}{\partial t} 
\right)^2\right]_{\ell=\ell_0} = 0  \label{equation2}
\end{eqnarray}\end{subequations}
Equation (23b) is the classical confinement 
condition (11) rewritten using the H-J formalism. The above 
conditions can be used not only to verify if  a given bulk metric
can lead to the required confinement but also to construct a new 
bulk metric with that property. In the former case one starts 
considering a known 5D metric and  solves eq. (20). Then, if 
eqs. (23a) and (23b) are fulfilled, it means that in the bulk 
with the considered 5D metric confinement is possible in the 
absence of non-gravitational forces. In the latter case, as we 
shall show in the following example, one can obtain  particular
5D metrics satisfying the H-J equation and the confinement 
constraints. In detail, assuming that (23a) is satisfied by a 
still unknown function $S_{t\ell}(t,\ell)$ and using (20), a 
simple way of satisfying (23b) is to put 
\begin{equation} n(t,\ell) = 
\dfrac{S_0\,\dfrac{dF(t)}{dt}}{\sqrt{m_5^2+\dfrac{C^2}{a^2(t,\ell)}}}
\end{equation}
where $S_0$ is a dimensionfull constant and $F(t)$ is an 
arbitrary dimensionless function of the time $t$. Then it will 
prove useful to split eq. (20) into 
\begin{subequations}\begin{eqnarray} 
\dfrac{1}{n}\,\left(\dfrac{\partial S_{t\ell}}{\partial t}\right)
= \sqrt{m_5^2 + \dfrac{C^2}{a^2}}\, 
\cosh{[\sqrt{\varepsilon}\,\beta(t,\ell)]} \label{equation3}\\
\dfrac{1}{\sqrt{\varepsilon}\,\Phi}\,\left(\dfrac{\partial 
S_{t\ell}}{\partial \ell}\right) = \sqrt{m_5^2 + 
\dfrac{C^2}{a^2}}\, \sinh{[\sqrt{\varepsilon}\,\beta(t,\ell)]}
\label{equation4}
\end{eqnarray}\end{subequations} 
where $\beta(t,\ell)$ is a  function to be determined and 
the Schwarz condition $\partial^2 S_{t\ell}/\partial \ell 
\partial t = \partial^2 S_{t\ell}/\partial t  \partial \ell$ has 
to be satisfied in the bulk. Clearly, in the case of massless 
particles one must require that $C^2 \neq 0$. Equation (23a) 
implies that $\sinh{[\sqrt{\varepsilon}\,\beta(t,\ell)]}=0$  on a
particular hypersurface $\Sigma_0$ for a value of $\ell$ equal to
a fixed $\ell_0$. Substituting (24) into (25a) we obtain
\begin{equation}
\left(\dfrac{\partial S_{t\ell}}{\partial t}\right) = 
S_0\,\dfrac{dF(t)}{dt}\,\cosh{[\sqrt{\varepsilon}\,\beta(t,\ell)]}
\end{equation}
The Schwarz condition is satisfied by the particularly simple 
choice where the function $\beta(t,\ell)$ is dependent only on 
$\ell$ in the form $\beta(\ell) = \kappa\, (\ell-\ell_0)$, 
$\kappa$ being a dimensionfull constant, and the fuction $\Phi$ 
is given by 
\begin{equation}
\Phi(t,\ell)=\dfrac{S_0\,F(t)\,\kappa}{\sqrt{m_5^2+\dfrac{C^2}{a^2(t,\ell)}}}
\end{equation}
From (25b) we have
\begin{equation}
\left(\dfrac{ \partial S_{t\ell}}{\partial \ell}\right) = 
S_0\,F(t)\,\kappa\,\sqrt{\varepsilon}
\sinh{[\sqrt{\varepsilon}\,\kappa\,(\ell-\ell_0)]}
\end{equation}
therefore we obtain the function $S_{t\ell}(t,\ell)$ as follows
\begin{equation}
S_{t\ell}(t,\ell) = 
S_0\,F(t)\,\cosh{[\sqrt{\varepsilon}\,\kappa\,(\ell-\ell_0)]} 
\end{equation} 
while the 5D line element (12) takes the form
\begin{widetext}\begin{equation}
ds_5^2= 
a^2(t,\ell)\,\left[d\sigma^2+\dfrac{S_0^2}{C^2+m_5^2\,a^2(t,\ell)}
\left(-\,\left(\dfrac{dF(t)}{dt}\right)^2\,dt^2+\varepsilon\, 
\kappa^2\, F^2(t)\,d\ell^2\right)\right] 
\end{equation}\end{widetext}
the function $a(t,\ell)$ and $F(t)$ remaining arbitrary in this 
particular solution. 
\section{C\lowercase{onclusions}} 
We combined the geodesic and  the Hamilton-Jacobi methods to 
select, in a cosmological scenario, 5D bulk metrics so that at 
least on one brane  of the foliation of a 5D manifold $M$  it is 
possible that test particles can be confined without the  
requirement of non-gravitational forces. Such a brane may 
represent our universe. Of course this does not exclude the 
possibility that confinement, if it must appear in any reasonable
theory with non-compact extra dimensions, can be obtained also by
means of  non-gravitational mechanisms. Here we would like to 
briefly discuss some practical conditions which, once fulfilled,
can ensure particle confinement. As a particularly simple 
example, a warped 5D metric of the form
\begin{equation}
ds_5^2 = e^{G(\ell)}\,\left[a^2(t)\,d\sigma^2 
-\,n^2(t)\,dt^2\right] + \varepsilon\,\Phi^2(t,\ell)\,d\ell^2 
\end{equation}
is apt to satisfy eq. (23b) provided that the warp factor  
$e^{G(\ell)}$ is such that 
$\left(d\,G(\ell)/d\,\ell\right)_{\ell=\ell_0} = 0$, while a 
function  $S_{t\ell}(t,\ell)$ of the form 
$S_{t\ell}(t,\ell) = S_0\,F(t)\, G(\ell)$, with the same 
$G(\ell)$ of the warp factor, is apt to satisfy eq. (23a). We 
note that eq. (20), whose solution  will give 
$S_{t\ell}(t,\ell)$, becomes more tractable if one isolates the 
effect of the extra dimension from the effects due to the motion 
in spacetime choosing the constant $C^2 = 0$. This simplification
would however not be possible in the massless case ($m_5^2 = 0$) 
if in eq. (20) also the term $\dfrac{1}{\varepsilon\,\Phi^2}\, 
\left(\dfrac{\partial S}{\partial \ell}\right)^2$ should become 
equal to zero. Finally let us see whether conditions for 
confinement are met for the well known 5D metric describing a 
class of cosmological models found by Ponce de Leon ${}^{27}$
\begin{equation}
ds_5^2 = \left(\dfrac{t}{L}\right)^{2/\alpha}
\,\left(\dfrac{\ell}{L}\right)^{2/(1-\alpha)}\,d\sigma^2-\,
\left(\dfrac{\ell}{L}\right)^2\,dt^2+\left(\dfrac{\alpha}{1-\,
\alpha}\right)^2\, \left(\dfrac{t}{L}\right)^2\,d\ell^2
\end{equation}
where $L$ is a constant length and $\alpha$ is a constant 
dimensionless parameter. The effects of the extra dimension as 
measured by an observer in 4D have been already examined in the 
background metric (32) but in a somewhat different context in 
${}^{19}$. We shall consider the following three  
possibilities:\pni 
\textbf{A)}\, $m_5^2> 0, C^2=0$.\pni
The solution to eq. (20) is
\begin{equation}
S_{t\ell}(t,\ell)= 
\dfrac{\alpha\,m_5}{\sqrt{2\alpha-\,1}\,L}\,t\,\ell
\end{equation}
Condition (23a) is not satisfied, so confinement is not possible.
The trajectory $\ell = \ell(\tau_4)$ as a function of the 4D 
proper time $\tau_4$  can be written as
\begin{equation}
\ell = L\,\left(\dfrac{\tau_4}{L}\right)^{-\,\left((\alpha 
-\,1)^2/\alpha^2\right)}
\end{equation}
From (17) evaluated along the trajectory (34) we obtain  in 
agreement with ${}^{19}$ the rest mass $m_4$ as 
\begin{equation}
m_4 =  \dfrac{\alpha\,m_5}{\sqrt{2\alpha-\,1}}
\end{equation} 
therefore $m_4$ takes on  different values in different 
cosmological ``eras'' marked by a particular choice for $\alpha$.
\pni 
\textbf{B)}\, $m_5^2=0, C^2=0$.\pni 
The solution to eq. (20) is
\begin{equation}
S_{t\ell}(t,\ell)= S_0\, 
\left(\dfrac{t}{L}\right)^A\,\left(\dfrac{\ell}{L}\right)^{A\,
\alpha/(\alpha -\,1)}
\end{equation}
where $A$ is a constant dimensionless parameter. As stated above,
in the massless case with $C^2=0$ one must not require that 
$\Phi^{-\,1}\,\left(\partial S_{t\ell}(t,\ell)/\partial 
\ell\right)$ vanishes so condition (23a), which otherwise
would be satisfied  at $\ell=0$,  can not be taken into account  
here and confinement is not possible. The trajectory $\ell =  
\ell(\tau_4)$ as a function of the 4D proper time $\tau_4$  can 
be written as  
\begin{equation} 
\ell = L\,\left(\dfrac{\tau_4}{\alpha \,L}\right)^{(1-\alpha)} 
\end{equation}
From (17) evaluated along the trajectory  (37) we obtain the rest
mass $m_4$ as 
\begin{equation}
m_4 = \dfrac{S_0\,A\,\alpha}{\tau_4}
\end{equation}
therefore, in agreement with ${}^{19}$, the variation of $m_4$ 
takes place in cosmological timescales.\pni
\textbf{C)}\, $m_5^2=0, C^2>0.$\pni
The solution to eq. (20) is
\begin{equation}
S_{t\ell}(t,\ell)=  
\dfrac{L\,C\,\alpha}{\sqrt{1-\,2\,\alpha}}\,\left(
\dfrac{t}{L}\right)^{(\alpha-\,1)/\alpha}\,\left(\dfrac{\ell}{L}
\right)^{\alpha/(\alpha-\,1)}
\end{equation}
The left-hand side of condition (23a) does not 
vanish at $\ell=0$ because here  $\alpha < 1/2$, so 
confinement is not possible. The trajectory $\ell = \ell(\tau_4)$
as a function of the 4D proper time $\tau_4$  can be written as  
\begin{equation} 
\ell = L\,\exp{\left(-\,\dfrac{\tau_4}{L}\right)} 
\end{equation}
From (17) evaluated along the trajectory  (40) we obtain the rest
mass $m_4$ as 
\begin{equation}
m_4 = 
\dfrac{C\,\alpha}{\sqrt{1-\,2\,\alpha}}\,\exp{\left[-\,\left(\dfrac{
\alpha^2+(1-\alpha)^2}{1-\alpha}\right)\,\left(\dfrac{\tau_4}{L}
\right)\right]}
\end{equation}
therefore again $m_4$ decreases on cosmological timescales. 
In all the three cases previously examined,  confinement
due only to classical gravitational effects is not possible.\pni
We conclude by  noting  that there are some 
questions which need to be investigated with regard to the 
stability of the confined trajectories, but this will be the 
subject of a future work. 
 
\end{document}